\documentclass[fleqn,usenatbib]{rasti}

\usepackage{newtxtext,newtxmath}

\usepackage[T1]{fontenc}

\DeclareRobustCommand{\VAN}[3]{#2}
\let\VANthebibliography\thebibliography
\def\thebibliography{\DeclareRobustCommand{\VAN}[3]{##3}\VANthebibliography}

\usepackage{graphicx}	
\usepackage{amsmath}	
\usepackage{natbib}

\title[]{Signature of exocomet transits observed with Ariel}

\author[N. Takács et al.]{
Nóra Takács$^{1,2},$\thanks{E-mail: takacs.nora@csfk.org}
Gyula M. Szabó$^{4,5}$,
Andreas Papageorgiou$^{6}$,
Lorenzo V. Mugnai$^{6}$,
\newauthor
Matt Griffin$^{6}$,
Giuseppe Morello$^{7,8}$,
Róbert Szabó$^{1,2,3}$,
and Csaba Kiss,$^{1,2}$
\\
$^{1}$Konkoly Observatory, HUN-REN Research Centre for Astronomy and Earth Sciences, MTA Center of Excellence, Konkoly Thege 15-17, H-1121 Budapest, Hungary\\
$^{2}$ELTE Eötvös Loránd University, Institute of Physics and Astronomy, Budapest, Hungary\\
$^{3}$MTA Centre of Excellence, Budapest, Konkoly Thege 15-17, H-1121\\
$^{4}$ELTE Eötvös Loránd University, Gothard Astrophysical Observatory, Szombathely, Szent Imre h. u. 112, 9700, Hungary\\
$^{5}$MTA-ELTE Lendület "Momentum" Milky Way Research Group, Szombathely, Szent Imre h. u. 112, 9700, Hungary\\
$^{6}$School of Physical, Chemical, and Environmental Sciences, Cardiff University, Queens Buildings, The Parade, Cardiff, CF24 3AA, UK\\
$^{7}$Instituto de Astrof\'isica de Andaluc\'ia (IAA-CSIC), Glorieta de  
la Astronom\'ia s/n, 18008 Granada, Spain\\
$^{8}$INAF- Palermo Astronomical Observatory, Piazza del Parlamento, 1,  
90134 Palermo, Italy\\
}

\date{Accepted XXX. Received YYY; in original form ZZZ}

\pubyear{\the\year{}}

\begin{document}
\label{firstpage}
\pagerange{\pageref{firstpage}--\pageref{lastpage}}
\maketitle

\begin{abstract}
We investigate the potential of the Ariel space telescope to detect and characterise exocomet transits through their wavelength-dependent photometric signatures. Building on a radiative-transfer framework, we model exocomet transits for three representative dust morphologies, three grain-size distributions, and four homogeneous dust compositions using synthetic photometric bands covering the Ariel wavelength range. The modelled events typically last 1--2~days and produce photometric variations of a few millimagnitudes. We find that the wavelength dependence of the transit signal provides stronger diagnostics of the dust properties than transit morphology alone. In particular, colour variations relative to the VISPhot band and their wavelength-dependent slopes separate the considered compositions into two broad groups: carbonaceous dust and graphite, and olivine and pyroxene. While the compositions within each group show similar behaviour, the two groups remain distinguishable across the dust morphologies and grain sizes considered. We further assess the detectability of these signals using the $\beta$~Pictoris spectral energy distribution and ArielRad simulations. Integrating over broad synthetic photometric bands substantially improves the signal-to-noise ratio, allowing millimagnitude-level signals to be detected with short integration times for sufficiently bright targets. Ariel's simultaneous optical-to-infrared coverage therefore offers the possibility not only of detecting exocomet transits, but also of characterising their chromatic signatures and constraining broad classes of dust properties. We also discuss the limitations of the modelling assumptions and the potential of coordinated observations with long-baseline photometric missions.
\end{abstract}




\section{Introduction}

Comets are remnants of the planet-formation process and preserve material inherited from the cold outer regions of protoplanetary disks. Their volatile and refractory components provide information on the chemical and physical conditions under which planetary systems formed, while their orbital evolution traces subsequent gravitational perturbations, planet migration, and the redistribution of planetesimals. Cometary impacts have also been proposed as a mechanism for transporting water and other volatiles into the inner regions of planetary systems, potentially influencing the composition and habitability of terrestrial planets \citep{Kral2018,Strom2020}. 

The first observational evidence for exocomets was obtained through variable circumstellar absorption features in the Ca II H and K lines of $\beta$ Pictoris \citep{Ferlet1987}. 
Long-term spectroscopic monitoring has revealed a large population of such bodies in the same system, including at least two families with different dynamical and physical properties \citep{Kiefer2014}. Similar variable absorption components have since been identified around other young stars and debris-disk hosts, although the number of securely established systems remains small \citep{Iglesias2025}. 

Exocomets can be detected photometrically when their dusty comae and tails pass in front of their host stars. Numerical calculations by \citet{1999A&AS..140...15L} predicted that such events would frequently produce asymmetric, approximately triangular transit profiles because the optically thickest part of the coma is followed or preceded by an extended dust tail. Candidate transiting exocomets were subsequently identified in Kepler photometry by \citet{Rappaport2018}, while TESS observations provided broadband detections of asymmetric events in the $\beta$ Pictoris system \citep{2019A&A...625L..13Z,2022NatSR..12.5855L}. 
These detections demonstrate that high-precision time-series photometry can probe the dust component of individual exocomets, complementing the gas traced by variable absorption lines.
Photometric exocomet identification nevertheless remains challenging. Individual events are generally shallow, non-periodic and unpredictable, and their shapes are not necessarily unique. Although an extended tail can produce a characteristic “shark-fin” profile, a coma-dominated comet may generate an approximately symmetric transit. Conversely, circumstellar dust clumps, disintegrating planets, stellar variability, instrumental systematics, and grazing eclipses can imitate some of the expected signatures. Transit morphology alone is therefore insufficient to establish a cometary origin in many cases. Independent information on the wavelength dependence of the event is needed both to reject false positives and to constrain the physical properties of the obscuring material.

Recent photometric searches are beginning to provide larger samples of candidate systems. \citet{2019MNRAS.482.5587K} identified an additional exocometary transit in the Kepler data, while the search of \citet{2025MNRAS.542.1486N} using TESS data recovered the previously known $\beta$~Pictoris event and several new exocomet candidates. \citet{2025A&A...704A.191D} identified new high-confidence exocometary transit candidates in addition to previously known events. These photometric searches could provide potential targets for follow-up observations with spectroscopic and spectrophotometric missions such as Ariel.

Spectroscopic observations provide complementary information on the composition of exocomets. In particular, \citet{2025A&A...697A..21V} measured the abundances of refractory ions in 29 individual $\beta$~Pictoris exocomets and found that overall refractory composition is broadly consistent with solar abundances, although individual species such as Ca II can be significantly depleted. More recently, \citet{2026A&A...707A..60V} used the excitation state of transiting gas to constrain the distances of three low-velocity exocomets in the $\beta$~Pictoris system to $0.88\pm 0.08$, $4.7\pm0.3$, and $1.52\pm 0.15$~au. These results demonstrate that spectroscopy can provide information on both the composition and physical environment of individual exocomets, while motivating complementary observations of their dust components.


In \citet{2024PASP..136h4401K} -- hereafter Paper I -- a radiative-transfer framework has been introduced for calculating exocomet transit light curves from arbitrary three-dimensional dust distributions. Three representative morphologies were considered —— a narrow tail, a fan-like tail, and a spherically symmetric coma —— together with carbon, graphite, pyroxene, and olivine dust. The characteristic grain sizes were set to 200~,nm, 650\, nm and 2\,$\mu$m to sample the total scattering cross section provided by \cite{2024MNRAS.527.3559P} and in accordance with Paper~I. Paper~I showed that small grains produce strongly wavelength-dependent transit depths, while larger grains generally produce flatter extinction signatures. It also demonstrated that asymmetric dust distributions can lead to wavelength-dependent shifts in the apparent transit minimum, and that multiband observations can distinguish broad classes of dust properties even when the transit morphology itself is degenerate.

The European Space Agency’s Ariel mission provides an opportunity to develop this approach from discrete-band photometry into time-resolved spectrophotometry. Ariel is designed to observe a sample of approximately 1000 known exoplanets using simultaneous visible and infrared measurements extending from approximately 0.5 to 7.8\,$\mu{\rm m}$ \citep{Tinetti2022}. Its broad instantaneous wavelength coverage is particularly useful for exocomet transits because the extinction and scattering signatures of different grain populations can be measured at the same epoch. The Ariel spectroscopic channels can therefore sample the wavelength dependence of the transit more densely than conventional broadband photometry, while the simultaneous optical measurements provide reference for the wavelength-dependent colour signal.

Ariel is not, however, a survey telescope optimised for discovering unpredictable exocomet transits. It will observe a predefined list of exoplanet systems at epochs chosen primarily from planetary transit and eclipse ephemerides. Exocomet detections will consequently be serendipitous unless known exocomet hosts, dippers, or young dusty systems are included in dedicated observations, either as core science targets or part of Ariel's Complementary Science Programme \citep{Tinetti2022}. 


The paper is structured as follows. In Sect.~\ref{Sect: Simulation setup}, we describe the simulation setup and the adopted dust models. Sect.~\ref{Modelled light curves and wavelength dependence} presents the modelled exocomet transit signatures and their wavelength dependence. In Sect.~\ref{Sect:Exocomet transit detectability in synthetic Ariel bands} we investigate the detectability of these signals with Ariel. Finally, Sect.~\ref{Sect:Conslusions} summarises our main findings and conclusions.


\section{Simulation setup}\label{Sect: Simulation setup}

\subsection{Radiative transfer simulations}\label{Sec: Radiative transfer simulations}

Radiative transfer calculations were performed with the RADMC-3D package \citep{radmc}, using version 2.0 on a Cartesian grid of $256^3$ cells. The physical size of the computational domain was adjusted to the adopted dust density distribution (Sect.~\ref{Sect: Dust density structures}), typically spanning $\sim0.05$\,au. The coordinate origin coincides with the center of the dust distribution, while the central star is placed on the $x$-axis at the heliocentric distance $r_h$, outside the computational domain. Throughout the light-curve simulations, we adopt a solar-like central star with $T_{\rm eff}=5780$\,K, $R_\star=R_\odot=6.96\times10^5$\,km, and $\log g=4.5$. This choice is used consistently for the calculation of the dust extinction, forward scattering, and transit light curves. The simulated transit durations are also consistent with the values expected for exocomet transit events reported in the literature \citep{2019A&A...625L..13Z, 2022A&A...660A..49P}. The modelled transit events in these models typically last approximately 1-2 days, depending on the exocomet morphology and viewing geometry. The relative magnitude variations are generally of the order of a few millimagnitudes, with typical values of approximately 2-3 mmag.

Dust optical properties were computed with the {\sl optool} package \citep{optool}. Each simulation assumes a homogeneous dust composition, while anisotropic scattering is treated with the Henyey--Greenstein approximation ({\sl scattering\_mode\_max=2}) using the scattering opacities and asymmetry parameters calculated by {\sl optool}.

The grain size distribution follows the standard power law \mbox{\(N(a)\propto a^{-q} \)},  with $q=3.5$ \citep{1969JGR....74.2531D}. We consider three grain size intervals: 0.1--0.3\,$\mu$m (143.8\,nm mean size), 0.3--1.0\,$\mu$m (439.5\,nm), and 1.0--3.0\,$\mu$m (1.44\,$\mu$m). These ranges cover the sub-micron to few-micron grains that contribute strongly to the wavelength-dependent extinction and scattering considered here. They are also within the range of dust-particle sizes observed in Solar System cometary comae, although cometary dust populations extend to substantially larger particles \citep{2019A&A...630A..26M}. Our selected intervals therefore do not represent the full cometary dust distribution, but rather the grain sizes most relevant for producing measurable optical-to-near-infrared colour signatures. Larger grains ($a\gtrsim10\,\mu$m) are not considered because their extinction becomes increasingly grey while their scattering is increasingly concentrated in the forward direction.


To isolate the influence of composition, each model is computed separately for four homogeneous materials: graphite \citep{Draine2003}, amorphous pyroxene (70\% Mg) \citep{Dorschner1995,Henning1996}, amorphous olivine (50\% Mg) \citep{Dorschner1995,Henning1996}, and amorphous carbon \citep{Zubko1996}. Although real cometary dust consists of mixtures of different species, treating each material independently allows us to quantify its individual effect on the transit signatures.

Extinction and forward-scattering light curves are calculated independently for a dust cloud orbiting a solar-type star on a circular orbit with $r_h=0.1$\,au. The orbit is assumed to be perfectly edge-on and the transit central, so the effects of orbital inclination are neglected \citep{1999A&AS..140...15L}. The transit is sampled by changing the observer's viewing angle in steps of $0.1^\circ$ during transit and $0.2^\circ$--$0.5^\circ$ outside transit, corresponding to the apparent motion of the dust cloud across the stellar disk. For each viewing geometry, RADMC-3D computes the optical-depth map, which is convolved with the limb-darkened stellar disk (Appendix~\ref{app:limb-darkening}) to obtain the extinction light curve.

\subsection{Dust density structures}\label{Sect: Dust density structures}

Dust density structures are modelled in the same way as in \citet{2024PASP..136h4401K}, where three dust density models were selected to span a range of tail morphologies and degrees of asymmetry, representing a narrow-tailed comet, a fan-shaped coma with a broad tail, and a nearly spherical dust cloud without a tail. Rather than employing a physically motivated dust production model \citep[e.g.][]{1999A&A...343..916L,1957Tell....9...92A,1968P&SS...16.1221W,1981A&A....98...45W}, we use analytical density distributions designed to reproduce the overall appearance of several well-known Solar System comets. This approach allows us to isolate the effect of the dust morphology on the resulting transit signatures.

\paragraph*{Narrow-tail model:}

This model is inspired by one of the fragments of Comet 73P/Schwassmann--Wachmann. It consists of a compact coma and a long, narrow dust tail, similar to the morphology reported by \citet{Reach2009}. The adopted density distribution is

\begin{equation}
 \rho = \Bigg\{
 \begin{array}{lr}
    w_1 \rho_0 e^{-r^2/\sigma_1^2}/r^2 + (1-w_1) \rho_0 e^{-(x^2+z^2)/\sigma_2^2} e^{-y^2/\sigma_f^2} & : y \ge 0 \\
    w_1 \rho_0 e^{-r^2/\sigma_1^2}/r^2 + (1-w_1) \rho_0 e^{-(x^2+z^2)/\sigma_2^2} e^{-|y|/\sigma_t} & : y < 0 ,
  \end{array}
\label{eq:73p}
\end{equation}

where $r^2=x^2+y^2+z^2$. The adopted parameters are
$w_1=0.2$, $\sigma_1=0.0015$\,au,
$\sigma_2=0.0025$\,au,
$\sigma_f=0.0025$\,au, and
$\sigma_t=0.010$\,au.
The finite extent of the tail reflects the limitations of our simplified model, which neglects both the continuous expansion of the tail and the variation in the star--particle distance along its length.

\paragraph*{Fan-tail model:}
This model resembles the morphology of Comet 29P/Schwassmann--Wachmann \citep{2008A&A...485..599T}. It consists of a dense central coma embedded in a diffuse envelope that extends preferentially in the $+y$ direction while declining rapidly in the opposite direction. The density is given by

\begin{equation}
 \rho = \Bigg\{
 \begin{array}{lr}
    w_1 \rho_0 r^{-2} e^{-r^2/\sigma_1^2} + (1-w_1) \rho_0 e^{-r^2/\sigma_2^2} & : y \ge 0 \\
    w_1 \rho_0 r^{-2} e^{-r^2/\sigma_1^2} + (1-w_1) \rho_0 e^{-(x^2+z^2)/\sigma_2^2} e^{-y^2/\sigma_t^2} & : y < 0 ,
  \end{array}
\label{eq:29p}
\end{equation}

with $w_1=0.5$,
$\sigma_1=0.003$\,au,
$\sigma_2=0.012$\,au, and
$\sigma_t=0.006$\,au.

\paragraph*{Tail-free model:} 
The third distribution is spherically symmetric and was chosen to mimic the appearance of Comet 17P/Holmes during its 2007 outburst \citep{2008ApJ...677L..63M,2008ApJ...684L..55M,2008LPI....39.1627T}. The density is represented by the sum of two Gaussian components,

\begin{equation}
    \rho = w_h \rho_0 e^{-r^2/\sigma_1^2} + (1-w_h) \rho_0 e^{-r^2/\sigma_2^2},
    \label{eq:holmes}
\end{equation}
\noindent
where $w_h=0.7$, $\sigma_1=0.002$\,au, and $\sigma_2=0.008$\,au.

For consistency with \citet{2024PASP..136h4401K}, the dust density
normalisation is chosen such that, for the $2\,\mu$m grain-size
distribution, the maximum optical depth of the extinction component
in the reference band is $\tau_R=10^{-4}$. This value was chosen
to correspond approximately to the maximum absorption depth observed
for exocomets in the $\beta$~Pictoris system
\citep{2022A&A...660A..49P,2022NatSR..12.5855L}. The resulting total dust masses of the
three density distributions are $0.84$--$2.29\times10^{11}$ kg for
the narrow-tail model, $4.65$--$12.6\times10^{11}$ kg for the
fan-tail model, and $2.96$--$6.98\times10^{11}$ kg for the tail-free
model, depending on the assumed composition
\citep{2024PASP..136h4401K}. The same total dust mass is retained
when changing the grain-size distribution, allowing the effect of
grain size on the transit signature to be investigated independently
of the total dust mass.


\subsection{Synthetic photometric bands}\label{Sect: synthetic photometric bands}

The synthetic photometric bands used in this work are listed in
Table~\ref{tab:synthetic_bands}. In addition to Ariel's dedicated
photometric channels, synthetic bands can be constructed by combining
spectral elements over broader wavelength intervals. This increases
the photometric precision at the expense of spectral resolution,
allowing the wavelength-dependent exocomet signals to be assessed
at different levels of spectral binning. The Ariel-specific band
definitions follow Gy. M. Szabó et al. (2026, RASTI, submitted).

\begin{table*}
    \centering
    \caption{Dedicated Ariel photometric channels and synthetic
    photometric bands used in this work. The central wavelengths
    and bandwidths define the wavelength intervals used to combine
    the spectroscopic measurements.}
    \label{tab:synthetic_bands}
    \begin{tabular}{l l c c c c}
        \hline
        Band & Instrument & Central wavelength &
        Bandwidth & Left edge & Right edge \\
        & & [$\mu$m] & [$\mu$m] & [$\mu$m] & [$\mu$m] \\
        \hline
        VISPhot & VISPhot & 0.550 & 0.100 & 0.500 & 0.600 \\
        FGS1 & FGS1 & 0.700 & 0.200 & 0.600 & 0.800 \\
        FGS2 & FGS2 & 0.950 & 0.300 & 0.800 & 1.100 \\
        J$_{\rm synth}$ & NIRSpec & 1.217 & 0.177 & 1.128 & 1.305 \\
        H$_{\rm synth}$ & NIRSpec & 1.630 & 0.238 & 1.511 & 1.749 \\
        K$_{\rm short,synth}$ & AIRS-CH0 & 2.195 & 0.392 & 1.999 & 2.391 \\
        K$_{\rm long,synth}$ & AIRS-CH0 & 2.736 & 0.784 & 2.344 & 3.128 \\
        IR1$_{\rm synth}$ & AIRS-CH0 & 3.547 & 0.837 & 3.128 & 3.965 \\
        IR2$_{\rm synth}$ & AIRS-CH1 & 4.543 & 0.891 & 4.097 & 4.988 \\
        IR3$_{\rm synth}$ & AIRS-CH1 & 5.656 & 1.657 & 4.827 & 6.484 \\
        \hline
    \end{tabular}
\end{table*}

The stellar limb-darkening coefficients adopted in the transit
simulations were calculated for a solar-like star and are provided
in Appendix~\ref{app:limb-darkening}.


\section{Results}\label{Sect:Results}

\subsection{Modelled light curves and wavelength dependence}\label{Modelled light curves and wavelength dependence}

\begin{figure*}
	\includegraphics[width=1.8\columnwidth]{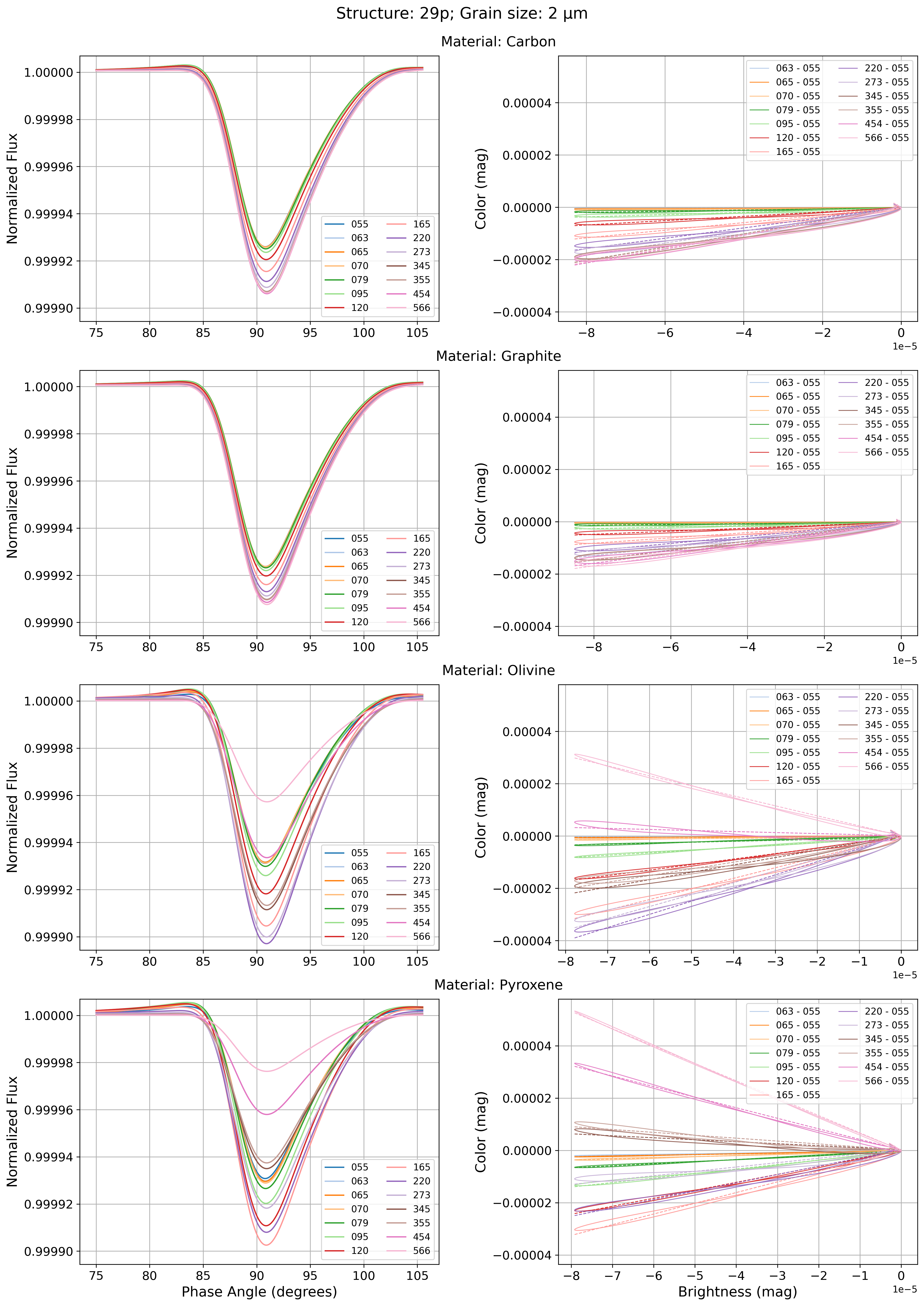}
    \caption{Left column: Modelled light curve for the narrow-tailed exocomet model with a 2 $\mu$m grain size. The four rows show different material compositions, and the colours of the solid lines represent different central wavelengths for the synthetic bands. Right column: The structure is the same as in the left column, but the solid lines show the colour difference between each central wavelength and the central wavelength of the VISPhot band. The dashed lines show the fitted slope parameter for each curve.} 
    \label{Fig: LightCurveAndColor_29p_c_gra_ol_pyr_2_055}
\end{figure*}

The exocomet light curves were simulated using the equations and method mentioned in Sections~\ref{Sec: Radiative transfer simulations} and ~\ref{Sect: Dust density structures}. 
An example of the results is presented in Fig.~\ref{Fig: LightCurveAndColor_29p_c_gra_ol_pyr_2_055}, along with the colour differences relative to the synthetic VISPhot central wavelength  (the simulated light curves and the corresponding figures can be found in the GitHub repository, linked in this work).

\begin{figure}
    \includegraphics[width=\columnwidth]{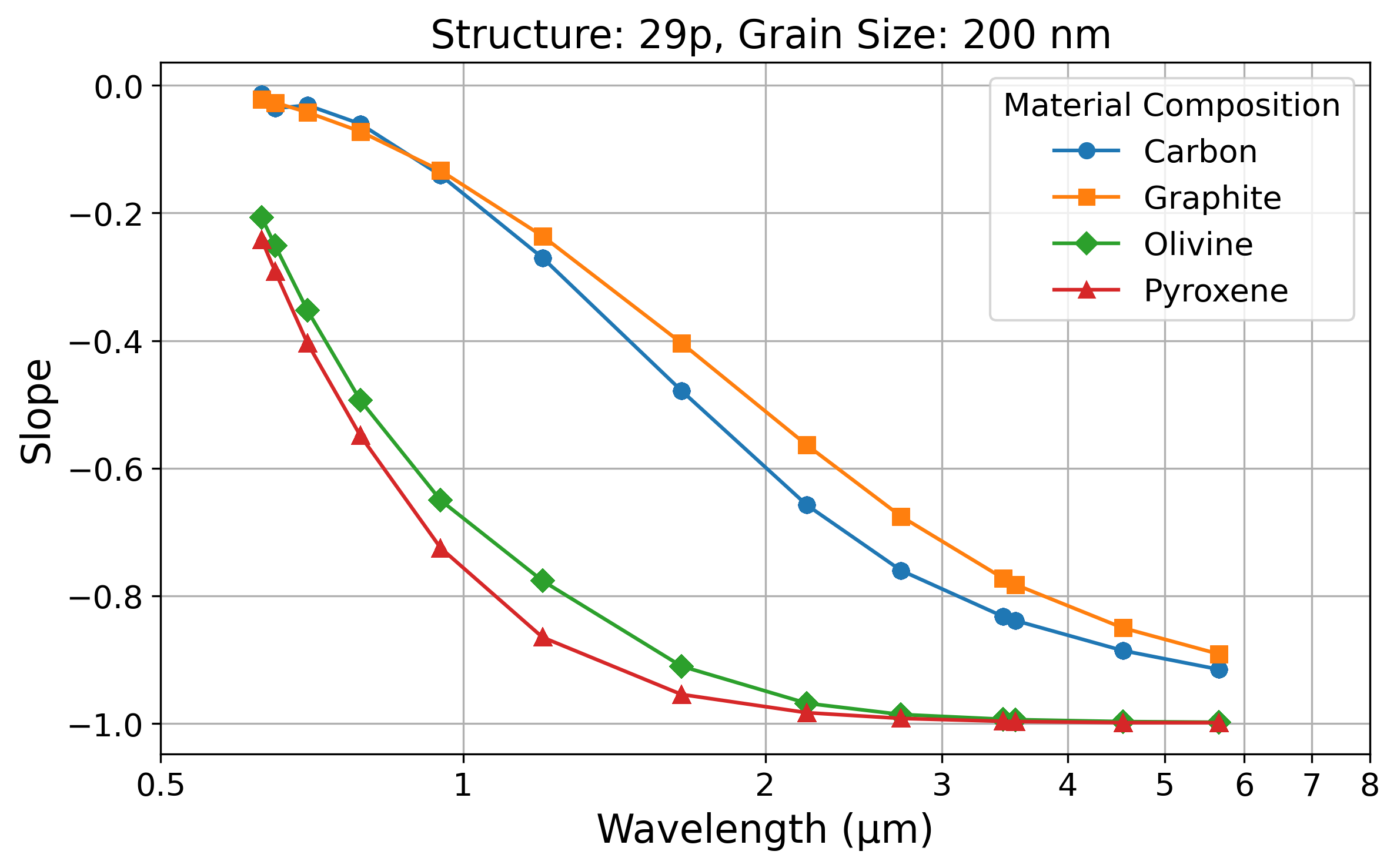}
    \includegraphics[width=\columnwidth]{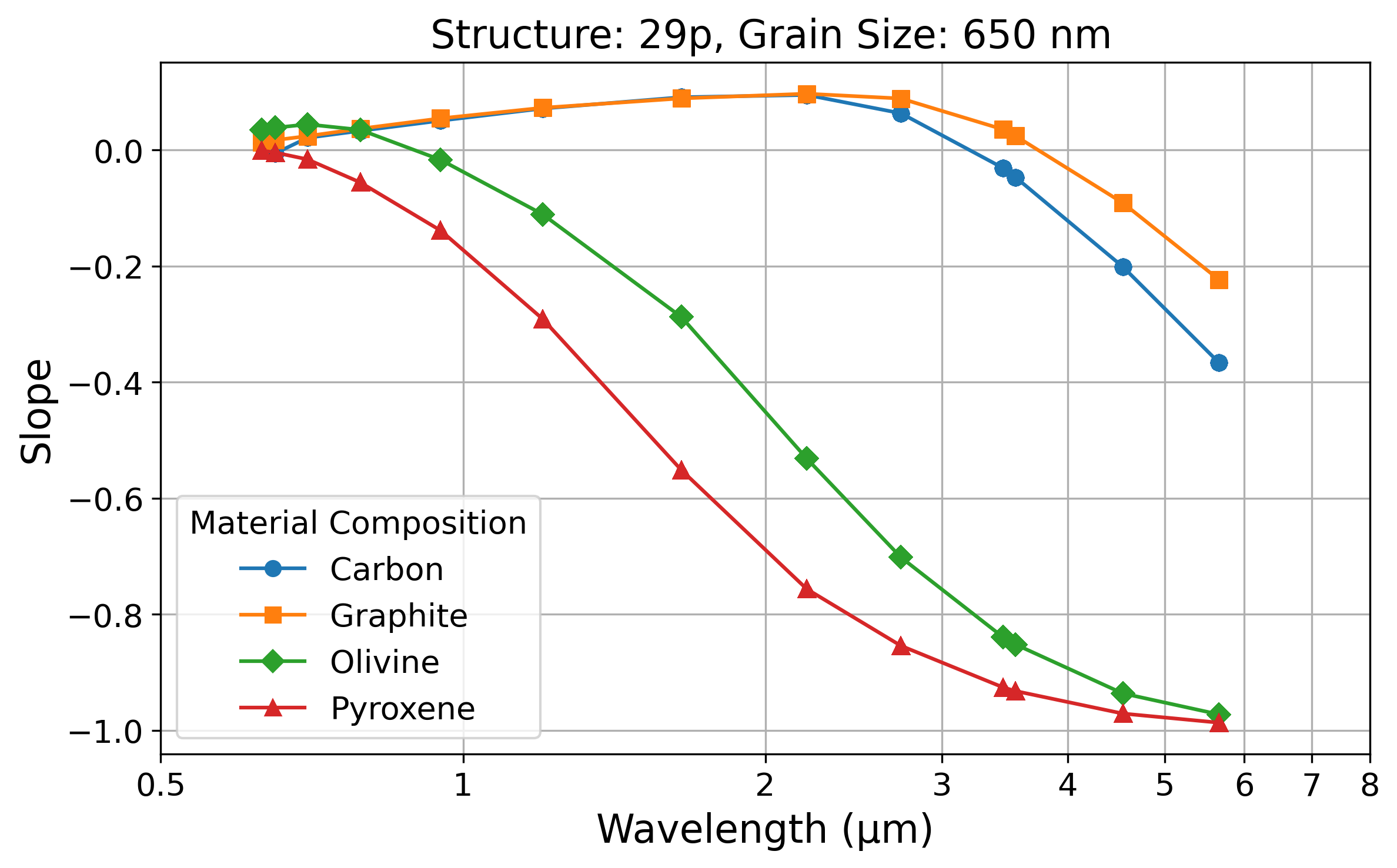}
	\includegraphics[width=\columnwidth]{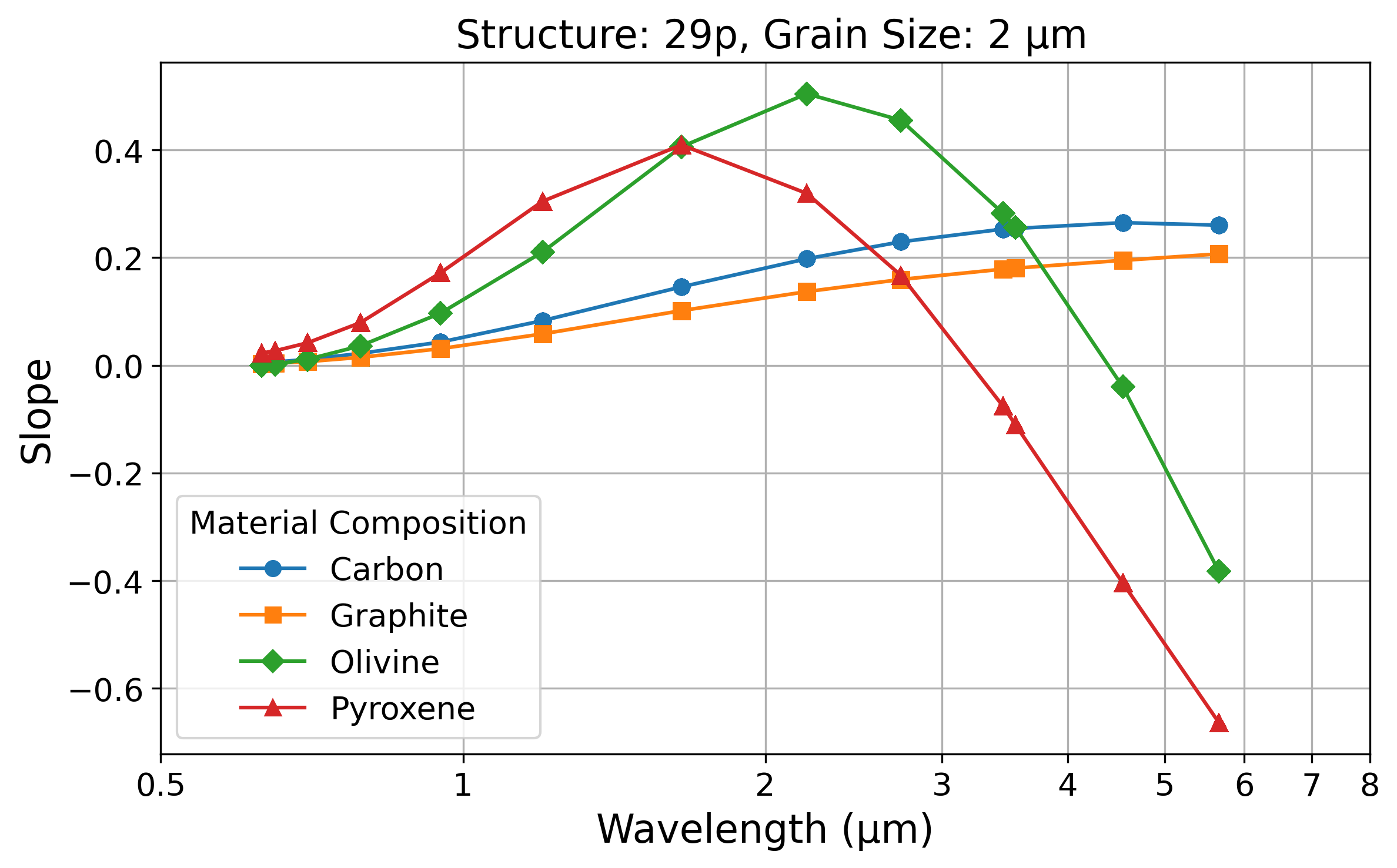}
    \caption{Slope parameters of the narrow-tailed exocomet model with grain sizes of 200 nm (top panel), 650 nm (middle panel) and 2 $\mu$m (bottom panel),  derived from the fit to the colour differences as a function of the central wavelengths. Each colour represents a different material.} 
    \label{Fig: slopes_29p_2}
\end{figure}

This shows the relative change in brightness between two wavelengths during the transit event and that the different dust compositions produce distinct wavelength-dependent signals. To be able to associate a measurable quantity to this colour difference, we carried out a linear fit to the colour-brightness curves as seen with dashed lines on the right panels of Fig.~\ref{Fig: LightCurveAndColor_29p_c_gra_ol_pyr_2_055}.
The value of the linear fit's slope describes the strength of the wavelength dependence: a greater deviation from zero corresponds to a greater change in the colour difference.
An example for the narrow-tailed exocomet model with a grain size of 2$\mu$m is shown in Fig.~\ref{Fig: slopes_29p_2}. These slope-wavelength relations suggest that the materials can be divided into two subgroups: carbonaceous dust and graphite, on the one hand, and olivine and pyroxene, on the other. The two materials within each group show similar behaviour throughout all the exocomet dust morphologies and grain sizes. 

Due to the overall independence from the actual morphology of the dust cloud, the extinction-- colour diagram, and especially the wavelength dependence of the slope parameter, is a good indicator of the dust properties: it has its internal degeneracies, but overall it is favourably specific to the dust type.  Most importantly, the two subgroups (namely carbon-graphite and olivine-pyroxene) can be clearly distinguished from one another.

\subsection{Exocomet transit detectability in synthetic Ariel bands}
\label{Sect:Exocomet transit detectability in synthetic Ariel bands}

To investigate the detectability of a dimming caused by an exocomet as observed with Ariel, we considered the spectral energy distribution (SED) of the $\beta$ Pictoris system as a representative bright exocomet-hosting system, in which several exocomets have already been detected. For this, we used IRTF/SpeX spectrum for the star's photosphere and Spitzer/IRS spectrum, containing infrared excess from the disc component, as published by \cite{2022ApJ...933...54L}.
Since there were still outliers in these spectra, with a gap between them, we cut out the parts with the outliers and fitted a Planck function to the photospheric spectra. We then filled in the gaps using the fitted function and connected the two spectra.

\begin{figure}
	\includegraphics[width=\columnwidth]{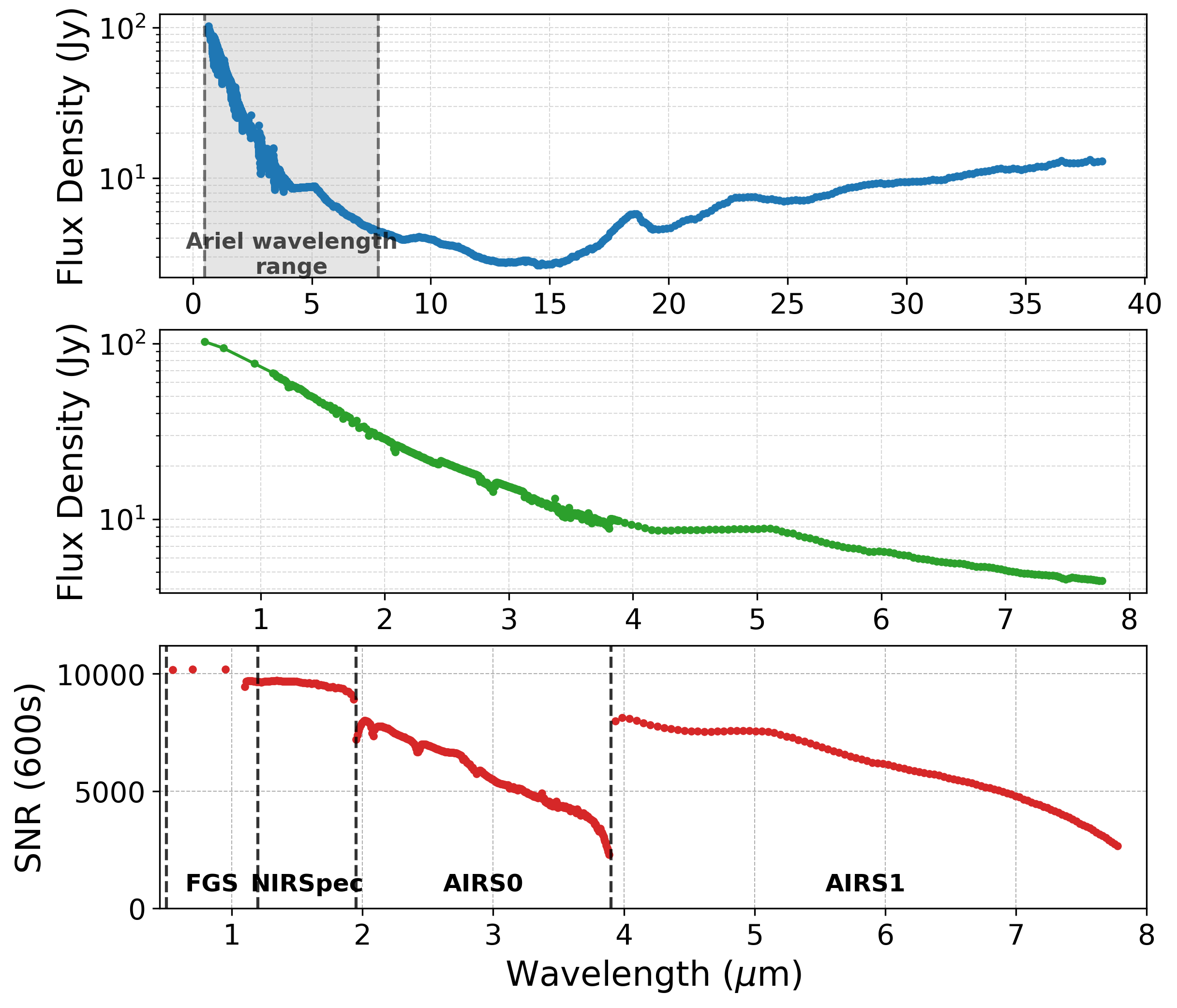}
    \caption{Spectral Energy Distribution for the $\beta$ Pictoris system. Upper panel: Original SED obtained from IRTF/SpeX and Spitzer/IRS \citep{2024yCat..19330054L}. Middle panel: Binned SED from ArielRad. Bottom panel: Simulation of the signal-to-noise ratio for the Ariel instruments from ArielRad.} 
    \label{fig:BetaPictorisArielRadThreePlots}
\end{figure}

Ariel sensitivity was characterised in terms of the noise-equivalent flux density (NEFD), defined here as the source flux density yielding a signal-to-noise ratio (SNR) of unity for an integration time of 0.5s. The wavelength- and source-flux-dependent NEFD was estimated using ArielRad \citep{2020ExA....50..303M}, adopting the current best estimates of the Ariel telescope and instrument characteristics and their associated noise contributions. For the input SED, the NEFD was then used to derive the expected SNR as a function of wavelength for a given integration time, at the native spectral resolution of each instrument. Higher SNR values can subsequently be obtained, at the expense of spectral resolution, by binning the simulated measurements over broader wavelength intervals. These simulations are intended for mission planning and preparatory studies only and should therefore be regarded as indicative, rather than predictive, of Ariel's final in-flight performance. The processed SED is shown in Fig.~\ref{fig:BetaPictorisArielRadThreePlots}, together with the corresponding ArielRad simulation, including the SED sampled at the native spectral resolution of the instruments and the associated SNR.

The simulated Ariel measurements were used to define the photometric precision expected in the synthetic bands. Since an exocomet obscures the stellar photosphere but not the circumstellar disc, the observed transit signal is diluted by the disc emission. We therefore calculated the photospheric fraction of the total flux in each band:

\begin{equation}
    f_{\mathrm{phot}} =
    \frac{F_{\mathrm{photosphere}}}
    {F_{\mathrm{photosphere}}+F_{\mathrm{disc}}},
\end{equation}
\noindent
where $F_{\mathrm{photosphere}}$ and $F_{\mathrm{disc}}$ are the photospheric and disc fluxes, respectively. For small dimming amplitudes, the observed transit depth can be approximated as

\begin{equation}
    \Delta m_{\mathrm{obs}}
    \approx f_{\mathrm{phot}}\Delta m_{\mathrm{star}},
\end{equation}

\noindent
where $\Delta m_{\mathrm{star}}$ is the intrinsic stellar dimming.

To estimate the signal-to-noise ratio of the synthetic photometric bands, we first assigned a weight to each spectral bin based on the number of photons it contributes. For a spectral bin centred at frequency $\nu_i$ with width $\Delta\nu_i$ and flux density $F_{\nu,i}$, the photon contribution is proportional to

\begin{equation}
    w_i=F_{\nu, i} \frac{\Delta \nu_i}{\nu_i}.
\end{equation}


\noindent
We then combined the signal-to-noise ratios of the individual spectral bins within each synthetic band using these photon weights:

\begin{equation}
    SNR_{synthetic} = \frac{\underset{i}{\sum}w_i}{\sqrt{\underset{i}{\sum}\left(\frac{w_i}{SNR_i}\right)^2}}
\end{equation}
\noindent
where $SNR_i$ is the signal-to-noise ratio obtained from the Ariel simulations in the specific spectral bins. 

We then calculated the expected transit detection S/N in each synthetic band using the combined band S/N and the fraction of the total flux contributed by the stellar photosphere. Since the exocomet obscures the star but not the circumstellar disc, the observed transit signal is diluted by the disc emission. The observed dimming in band $i$ is therefore

\begin{equation}
    \Delta m_{\mathrm{obs},i} = f_{\mathrm{phot},i}\Delta m_{\mathrm{intrinsic}},
\end{equation}

\noindent
where $f_{\mathrm{phot},i}$ is the photospheric fraction of the total flux and $\Delta m_{\mathrm{intrinsic}}$ is the intrinsic stellar dimming. We estimated the corresponding detection S/N as

\begin{equation}
    SNR_{\mathrm{transit},i}=\frac{\Delta m_{\mathrm{obs},i}SNR_{\mathrm{synthetic},i}}{1.0857}.
\end{equation}

\begin{figure}
	\includegraphics[width=\columnwidth]{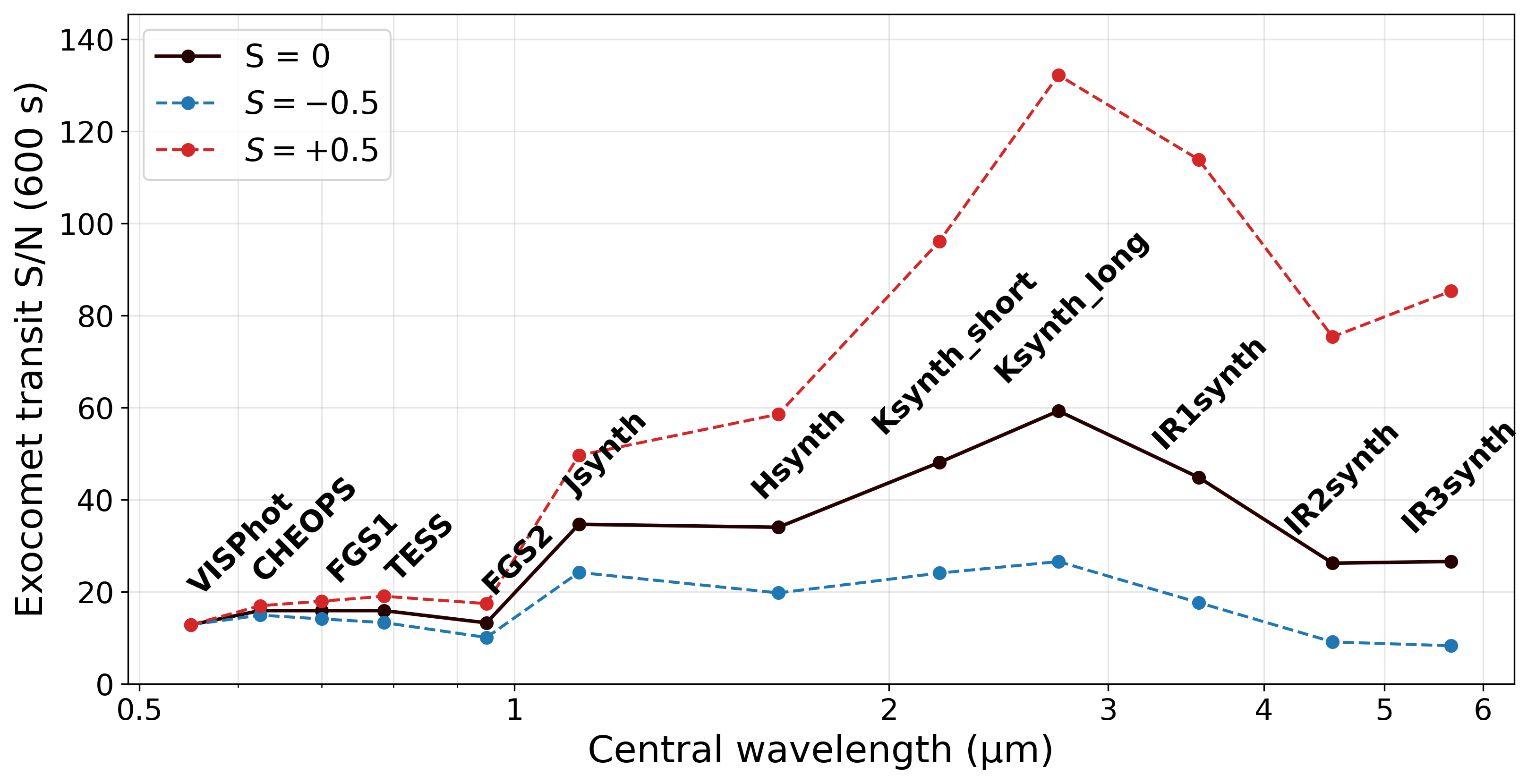}
    \caption{Expected exocomet transit detection S/N as a function of central wavelength for 600\,s integration, assuming an intrinstic dimming of 1\,mmag for the $\beta$~Pictoris SED. The solid curve ($S=0$) assumes a wavelength-independent intrinsic dimming. The dashed curves ($S=\pm0.5$) show illustrative wavelength-dependent signal scaling relative to the $S=0$ case.}
    \label{fig:ExocometTransit_SNR_600s_S0}
\end{figure}

Fig.~\ref{fig:ExocometTransit_SNR_600s_S0} shows the expected transit detection S/N as a function of wavelength for a 600\,s integration, assuming an intrinsic dimming of 1\,mmag. The solid curve ($S=0$) represents a wavelength-independent dimming, while the dashed curves illustrate how the expected detection S/N changes under alternative wavelength-dependent signal scalings corresponding to $S=-0.5$ and $S=+0.5$. These curves illustrate hypothetical wavelength-dependent signal scalings. These are theoretical examples, as no specific dust model is expected to maintain a constant colour slope across all synthetic wavelength bands.

The transit detection S/N scales approximately with the square root of the integration time under the assumed noise model. Therefore, the values can be rescaled for other integration times as $SNR(t)=SNR(600\,\mathrm{s})\sqrt{t/(600\,\mathrm{s})}$. Since the modelled exocomet transits typically last on the order of days, longer integrations may improve the detection significance, subject to the desired temporal resolution.

\subsection{Observational synergies}\label{Sect:Observational synergies}

During its mission, TESS has obtained observations using different cadences. In Cycles 1-2, TESS observed with a 30\,min cadence that was shortened to 10 minutes during Cycles 3-4 and to 200s from Cycle 5 onwards \citep{TESS_SDPDD_2020}. The heatmaps show that the few-millimagnitude signals can be detected at several $\sigma$ significance even at shorter integration times, with the detection significance increasing substantially for the longer TESS cadences. For example, a 2-3 mmag signal can reach approximately 5$\sigma$ significance in several of the considered bands at integration times of a few minutes.

The synthetic Ariel bands generally provide improved sensitivity compared with the TESS bands, particularly for the broader bands, allowing comparable detection significances to be reached with shorter integration times or for smaller dimming amplitudes. In addition to this improved photometric precision, Ariel provides observations over a much broader wavelength range, which allows the wavelength-dependent colour signal to be investigated and can therefore provide information on the composition of the exocomet.

Another possibility for synergies and coordinated observations will be the observations obtained by PLATO \citep{2014ExA....38..249R}. The PLATO observing strategy is designed to provide exceptionally long temporal coverage of selected fields, with the first long-pointing field, LOPS2, scheduled to be monitored continuously for at least two years \citep{2025A&A...694A.313N}. This extended baseline is particularly advantageous for coordinated observations with Ariel, since the temporal constraint on an Ariel observation need not be set by the joint observability windows of all instruments involved in the synergy, but primarily by Ariel's own observability of the target. Should PLATO, in favourable cases, discover recurrent transit-like dips associated with dusty bodies, such discoveries would provide particularly valuable targets for intensive follow-up observations aimed at characterising their dust properties and potentially revealing a previously unknown class of solar system objects.






\section{Conclusions}\label{Sect:Conslusions}

We have investigated the potential of Ariel to detect and characterise exocomet transits through radiative-transfer simulations spanning three representative dust morphologies, three grain-size distributions, and four homogeneous dust compositions. Our results show that wavelength-dependent colour variations provide a more robust diagnostic of dust properties than transit morphology alone. In particular, the spectral dependence of the colour–extinction relation separates carbonaceous dust and graphite from olivine and pyroxene across the morphologies and grain-size ranges considered. However, the similarities between materials within each group limit the possibility of unique compositional identification.

The detectability analysis, based on the spectral energy distribution of $\beta$~Pictoris and ArielRad simulations, indicates that millimagnitude-level dimming can be measured at significant levels for sufficiently bright targets. Broad synthetic passbands improve photometric precision, while simultaneous optical-to-infrared measurements retain the wavelength-dependent information needed to distinguish broad dust families.

Ariel is not designed to survey for unpredictable exocomet transits, so its contribution will depend on the availability of suitable targets and observing opportunities. Coordinated monitoring with missions such as TESS and PLATO could identify candidate events or constrain their recurrence, enabling Ariel follow-up to test their chromatic signatures and characterise their dust properties. This approach would complement the discovery potential of long-baseline photometry with Ariel's simultaneous wavelength coverage.

The present calculations are deliberately idealised: they assume homogeneous dust compositions, analytical density distributions, a central edge-on transit at a fixed star–comet distance, and a solar-type illuminating star. The predicted sensitivities are also based on pre-launch estimates of Ariel’s performance. Consequently, the numerical detection limits should be treated as indicative rather than universal. Future work should extend the models to mixed compositions, physically evolving grain populations, a broader range of stellar and orbital parameters, and a retrieval framework that propagates instrumental and astrophysical covariance.

\section*{Acknowledgements}

The research leading to these results received funding from the PRODEX programme of the European Space Agency under the contract PEA~4000149210. N.T. has been supported by the EKÖP-25 University Research Scholarship Program of the Ministry for Culture and Innovation, from the source of the National Research, Development and Innovation Fund, Hungary. 
G.M. acknowledges financial support from the Severo Ochoa grant CEX2021-001131-S and from the Ramón y Cajal grant RYC2022-037854-I  funded by MICIU/AEI/1144 10.13039/501100011033 and FSE+.
Gy.M.Sz. and R.Sz. thank the SNN-147362 and the ADVANCED-153410 of the National Research, Development and Innovation Office (NKFIH, Hungary), and Gy.M.Sz. thanks to the ESA PRODEX Experiment Agreements No. 4000137122 and No. 4000149203. R.Sz. also acknowledges the LP2025-14/2025 Lendület grant of the Hungarian Academy of Sciences.

\section*{Data Availability}\label{Sect: Data Availability}

The simulated light curves obtained in this study and the figures generated from them are available at \href{https://github.com/nora-takacs-97/exocomets_with_ariel}{github.com} (https://github.com/nora-takacs-97/exocomets\_with\_ariel).

\section*{Conflict of Interest}\label{Sect:CoI}
The authors declare no conflict of interest.



\bibliographystyle{rasti}
\bibliography{refs} 

@ARTICLE{2024PASP..136h4401K,
       author = {{K{\'a}lm{\'a}n}, Szil{\'a}rd and {Szab{\'o}}, Gyula M. and {Kiss}, Csaba},
        title = "{Exocomet Models in Transit: Light Curve Morphology in the Optical{\textemdash}Near Infrared Wavelength Range}",
      journal = {\pasp},
         year = 2024,
        month = aug,
       volume = {136},
       number = {8},
          eid = {084401},
        pages = {084401},
          doi = {10.1088/1538-3873/ad4fe3},
archivePrefix = {arXiv},
       eprint = {2405.13663},
 primaryClass = {astro-ph.EP},
       adsurl = {https://ui.adsabs.harvard.edu/abs/2024PASP..136h4401K}
}

@ARTICLE{2020AJ....159...75M,
       author = {{Morello}, G. and {Claret}, A. and {Martin-Lagarde}, M. and {Cossou}, C. and {Tsiaras}, A. and {Lagage}, P.-O.},
        title = "{The ExoTETHyS Package: Tools for Exoplanetary Transits around Host Stars}",
      journal = {\aj},
         year = 2020,
        month = feb,
       volume = {159},
       number = {2},
          eid = {75},
        pages = {75},
          doi = {10.3847/1538-3881/ab63dc},
archivePrefix = {arXiv},
       eprint = {1908.09599},
 primaryClass = {astro-ph.EP},
       adsurl = {https://ui.adsabs.harvard.edu/abs/2020AJ....159...75M}
}

@ARTICLE{2020JOSS....5.1834M,
       author = {{Morello}, Giuseppe and {Claret}, Antonio and {Martin-Lagarde}, Marine and {Cossou}, Christophe and {Tsiara}, Angelos and {Lagage}, Pierre-Olivier},
        title = "{ExoTETHyS: Tools for Exoplanetary Transits around host stars}",
      journal = {The Journal of Open Source Software},
         year = 2020,
        month = feb,
       volume = {5},
       number = {46},
          eid = {1834},
        pages = {1834},
          doi = {10.21105/joss.01834},
       adsurl = {https://ui.adsabs.harvard.edu/abs/2020JOSS....5.1834M}
}

@ARTICLE{2022ExA....53..533M,
       author = {{Morello}, Giuseppe and {Danielski}, Camilla and {Sarkar}, Subhajit},
        title = "{The Ariel 0.6 - 7.8 {\ensuremath{\mu}}m stellar limb-darkening coefficients}",
      journal = {Experimental Astronomy},
         year = 2022,
        month = apr,
       volume = {53},
       number = {2},
        pages = {533-545},
          doi = {10.1007/s10686-021-09740-w},
archivePrefix = {arXiv},
       eprint = {2104.08898},
 primaryClass = {astro-ph.SR},
       adsurl = {https://ui.adsabs.harvard.edu/abs/2022ExA....53..533M}
}

@ARTICLE{2013A&A...553A...6H,
       author = {{Husser}, T.-O. and {Wende-von Berg}, S. and {Dreizler}, S. and {Homeier}, D. and {Reiners}, A. and {Barman}, T. and {Hauschildt}, P.~H.},
        title = "{A new extensive library of PHOENIX stellar atmospheres and synthetic spectra}",
      journal = {\aap},
         year = 2013,
        month = may,
       volume = {553},
          eid = {A6},
        pages = {A6},
          doi = {10.1051/0004-6361/201219058},
archivePrefix = {arXiv},
       eprint = {1303.5632},
 primaryClass = {astro-ph.SR},
       adsurl = {https://ui.adsabs.harvard.edu/abs/2013A&A...553A...6H}
}

@ARTICLE{2012A&A...546A..14C,
       author = {{Claret}, A. and {Hauschildt}, P.~H. and {Witte}, S.},
        title = "{New limb-darkening coefficients for PHOENIX/1D model atmospheres. I. Calculations for 1500 K {\ensuremath{\leq}} T$_{eff}$ {\ensuremath{\leq}} 4800 K Kepler, CoRot, Spitzer, uvby, UBVRIJHK, Sloan, and 2MASS photometric systems}",
      journal = {\aap},
         year = 2012,
        month = oct,
       volume = {546},
          eid = {A14},
        pages = {A14},
          doi = {10.1051/0004-6361/201219849},
       adsurl = {https://ui.adsabs.harvard.edu/abs/2012A&A...546A..14C}
}

@ARTICLE{2013A&A...552A..16C,
       author = {{Claret}, A. and {Hauschildt}, P.~H. and {Witte}, S.},
        title = "{New limb-darkening coefficients for Phoenix/1d model atmospheres. II. Calculations for 5000 K {\ensuremath{\leq}} T$_{eff}$ {\ensuremath{\leq}} 10 000 K Kepler, CoRot, Spitzer, uvby, UBVRIJHK, Sloan, and 2MASS photometric systems}",
      journal = {\aap},
         year = 2013,
        month = apr,
       volume = {552},
          eid = {A16},
        pages = {A16},
          doi = {10.1051/0004-6361/201220942},
       adsurl = {https://ui.adsabs.harvard.edu/abs/2013A&A...552A..16C}
}

@ARTICLE{2000A&A...363.1081C,
       author = {{Claret}, A.},
        title = "{A new non-linear limb-darkening law for LTE stellar atmosphere models. Calculations for -5.0 <= log[M/H] <= +1, 2000 K <= T$_{eff}$ <= 50000 K at several surface gravities}",
      journal = {\aap},
         year = 2000,
        month = nov,
       volume = {363},
        pages = {1081-1190},
       adsurl = {https://ui.adsabs.harvard.edu/abs/2000A&A...363.1081C}
}

@ARTICLE{1969JGR....74.2531D,
       author = {{Dohnanyi}, J.~S.},
        title = "{Collisional Model of Asteroids and Their Debris}",
      journal = {\jgr},
         year = 1969,
        month = may,
       volume = {74},
        pages = {2531-2554},
          doi = {10.1029/JB074i010p02531},
       adsurl = {https://ui.adsabs.harvard.edu/abs/1969JGR....74.2531D}
}

@ARTICLE{2024MNRAS.527.3559P,
       author = {{Pawellek}, Nicole and {Mo{\'o}r}, Attila and {Kirchschlager}, Florian and {Milli}, Julien and {K{\'o}sp{\'a}l}, {\'A}gnes and {{\'A}brah{\'a}m}, P{\'e}ter and {Marino}, Sebastian and {Wyatt}, Mark and {Rebollido}, Isabel and {Hughes}, A. Meredith and {Cantalloube}, Faustine and {Henning}, Thomas},
        title = "{The debris disc of HD 131488: bringing together thermal emission and scattered light}",
      journal = {\mnras},
         year = 2024,
        month = jan,
       volume = {527},
       number = {2},
        pages = {3559-3584},
          doi = {10.1093/mnras/stad3455},
archivePrefix = {arXiv},
       eprint = {2311.03272},
 primaryClass = {astro-ph.EP},
       adsurl = {https://ui.adsabs.harvard.edu/abs/2024MNRAS.527.3559P}
}

@ARTICLE{2019MNRAS.482.5587K,
       author = {{Kennedy}, Grant M. and {Hope}, Greg and {Hodgkin}, Simon T. and {Wyatt}, Mark C.},
        title = "{An automated search for transiting exocomets}",
      journal = {\mnras},
         year = 2019,
        month = feb,
       volume = {482},
       number = {4},
        pages = {5587-5596},
          doi = {10.1093/mnras/sty3049},
archivePrefix = {arXiv},
       eprint = {1811.03102},
 primaryClass = {astro-ph.EP},
       adsurl = {https://ui.adsabs.harvard.edu/abs/2019MNRAS.482.5587K}
}

@ARTICLE{2022ApJ...933...54L,
       author = {{Lu}, Cicero X. and {Chen}, Christine H. and {Sargent}, B.~A. and {Watson}, Dan M. and {Lisse}, Carey M. and {Green}, Joel D. and {Sitko}, Michael L. and {Mittal}, Tushar and {Lebouteiller}, V. and {Sloan}, G.~C. and {Rebollido}, Isabel and {Hines}, Dean C. and {Girard}, Julien H. and {Werner}, Michael W. and {Stapelfeldt}, Karl R. and {Wu}, Winston and {Worthen}, Kadin},
        title = "{Trends in Silicates in the {\ensuremath{\beta}} Pictoris Disk}",
      journal = {\apj},
         year = 2022,
        month = jul,
       volume = {933},
       number = {1},
          eid = {54},
        pages = {54},
          doi = {10.3847/1538-4357/ac70d1},
archivePrefix = {arXiv},
       eprint = {2205.09138},
 primaryClass = {astro-ph.EP},
       adsurl = {https://ui.adsabs.harvard.edu/abs/2022ApJ...933...54L}
}

@ARTICLE{2025A&A...694A.313N,
       author = {{Nascimbeni}, V. and {Piotto}, G. and {Cabrera}, J. and {Montalto}, M. and {Marinoni}, S. and {Marrese}, P.~M. and {Aerts}, C. and {Altavilla}, G. and {Benatti}, S. and {B{\"o}rner}, A. and {Deleuil}, M. and {Desidera}, S. and {Gizon}, L. and {Goupil}, M.~J. and {Granata}, V. and {Heras}, A.~M. and {Magrin}, D. and {Malavolta}, L. and {Mas-Hesse}, J.~M. and {Osborn}, H.~P. and {Pagano}, I. and {Paproth}, C. and {Pollacco}, D. and {Prisinzano}, L. and {Ragazzoni}, R. and {Ramsay}, G. and {Rauer}, H. and {Tkachenko}, A. and {Udry}, S.},
        title = "{The PLATO field selection process: II. Characterization of LOPS2, the first long-pointing field}",
      journal = {\aap},
         year = 2025,
        month = feb,
       volume = {694},
          eid = {A313},
        pages = {A313},
          doi = {10.1051/0004-6361/202452325},
archivePrefix = {arXiv},
       eprint = {2501.07687},
 primaryClass = {astro-ph.EP},
       adsurl = {https://ui.adsabs.harvard.edu/abs/2025A&A...694A.313N}
}

@ARTICLE{2008A&A...485..599T,
       author = {{Trigo-Rodr{\'\i}guez}, J.~M. and {Garc{\'\i}a-Melendo}, E. and {Davidsson}, B.~J.~R. and {S{\'a}nchez}, A. and {Rodr{\'\i}guez}, D. and {Lacruz}, J. and {de Los Reyes}, J.~A. and {Pastor}, S.},
        title = "{Outburst activity in comets. I. Continuous monitoring of comet 29P/Schwassmann-Wachmann 1}",
      journal = {\aap},
         year = 2008,
        month = jul,
       volume = {485},
       number = {2},
        pages = {599-606},
          doi = {10.1051/0004-6361:20078666},
       adsurl = {https://ui.adsabs.harvard.edu/abs/2008A&A...485..599T}
}

@ARTICLE{2014ExA....38..249R,
       author = {{Rauer}, H. and {Catala}, C. and {Aerts}, C. and {Appourchaux}, T. and {Benz}, W. and {Brandeker}, A. and {Christensen-Dalsgaard}, J. and {Deleuil}, M. and {Gizon}, L. and {Goupil}, M. -J. and {G{\"u}del}, M. and {Janot-Pacheco}, E. and {Mas-Hesse}, M. and {Pagano}, I. and {Piotto}, G. and {Pollacco}, D. and {Santos}, {\.{C}}. and {Smith}, A. and {Su{\'a}rez}, J. -C. and {Szab{\'o}}, R. and {Udry}, S. and {Adibekyan}, V. and {Alibert}, Y. and {Almenara}, J. -M. and {Amaro-Seoane}, P. and {Eiff}, M. Ammler-von and {Asplund}, M. and {Antonello}, E. and {Barnes}, S. and {Baudin}, F. and {Belkacem}, K. and {Bergemann}, M. and {Bihain}, G. and {Birch}, A.~C. and {Bonfils}, X. and {Boisse}, I. and {Bonomo}, A.~S. and {Borsa}, F. and {Brand{\~a}o}, I.~M. and {Brocato}, E. and {Brun}, S. and {Burleigh}, M. and {Burston}, R. and {Cabrera}, J. and {Cassisi}, S. and {Chaplin}, W. and {Charpinet}, S. and {Chiappini}, C. and {Church}, R.~P. and {Csizmadia}, Sz. and {Cunha}, M. and {Damasso}, M. and {Davies}, M.~B. and {Deeg}, H.~J. and {D{\'\i}az}, R.~F. and {Dreizler}, S. and {Dreyer}, C. and {Eggenberger}, P. and {Ehrenreich}, D. and {Eigm{\"u}ller}, P. and {Erikson}, A. and {Farmer}, R. and {Feltzing}, S. and {de Oliveira Fialho}, F. and {Figueira}, P. and {Forveille}, T. and {Fridlund}, M. and {Garc{\'\i}a}, R.~A. and {Giommi}, P. and {Giuffrida}, G. and {Godolt}, M. and {Gomes da Silva}, J. and {Granzer}, T. and {Grenfell}, J.~L. and {Grotsch-Noels}, A. and {G{\"u}nther}, E. and {Haswell}, C.~A. and {Hatzes}, A.~P. and {H{\'e}brard}, G. and {Hekker}, S. and {Helled}, R. and {Heng}, K. and {Jenkins}, J.~M. and {Johansen}, A. and {Khodachenko}, M.~L. and {Kislyakova}, K.~G. and {Kley}, W. and {Kolb}, U. and {Krivova}, N. and {Kupka}, F. and {Lammer}, H. and {Lanza}, A.~F. and {Lebreton}, Y. and {Magrin}, D. and {Marcos-Arenal}, P. and {Marrese}, P.~M. and {Marques}, J.~P. and {Martins}, J. and {Mathis}, S. and {Mathur}, S. and {Messina}, S. and {Miglio}, A. and {Montalban}, J. and {Montalto}, M. and {Monteiro}, M.~J.~P.~F.~G. and {Moradi}, H. and {Moravveji}, E. and {Mordasini}, C. and {Morel}, T. and {Mortier}, A. and {Nascimbeni}, V. and {Nelson}, R.~P. and {Nielsen}, M.~B. and {Noack}, L. and {Norton}, A.~J. and {Ofir}, A. and {Oshagh}, M. and {Ouazzani}, R. -M. and {P{\'a}pics}, P. and {Parro}, V.~C. and {Petit}, P. and {Plez}, B. and {Poretti}, E. and {Quirrenbach}, A. and {Ragazzoni}, R. and {Raimondo}, G. and {Rainer}, M. and {Reese}, D.~R. and {Redmer}, R. and {Reffert}, S. and {Rojas-Ayala}, B. and {Roxburgh}, I.~W. and {Salmon}, S. and {Santerne}, A. and {Schneider}, J. and {Schou}, J. and {Schuh}, S. and {Schunker}, H. and {Silva-Valio}, A. and {Silvotti}, R. and {Skillen}, I. and {Snellen}, I. and {Sohl}, F. and {Sousa}, S.~G. and {Sozzetti}, A. and {Stello}, D. and {Strassmeier}, K.~G. and {{\v{S}}vanda}, M. and {Szab{\'o}}, Gy. M. and {Tkachenko}, A. and {Valencia}, D. and {Van Grootel}, V. and {Vauclair}, S.~D. and {Ventura}, P. and {Wagner}, F.~W. and {Walton}, N.~A. and {Weingrill}, J. and {Werner}, S.~C. and {Wheatley}, P.~J. and {Zwintz}, K.},
        title = "{The PLATO 2.0 mission}",
      journal = {Experimental Astronomy},
         year = 2014,
        month = nov,
       volume = {38},
       number = {1-2},
        pages = {249-330},
          doi = {10.1007/s10686-014-9383-4},
archivePrefix = {arXiv},
       eprint = {1310.0696},
 primaryClass = {astro-ph.EP},
       adsurl = {https://ui.adsabs.harvard.edu/abs/2014ExA....38..249R}
}

@ARTICLE{2008ApJ...677L..63M,
       author = {{Moreno}, F. and {Ortiz}, J.~L. and {Santos-Sanz}, P. and {Morales}, N. and {Vidal-N{\'u}{\~n}ez}, M.~J. and {Lara}, L.~M. and {Guti{\'e}rrez}, P.~J.},
        title = "{A Model of the Early Evolution of the 2007 Outburst of Comet 17P/Holmes}",
      journal = {\apjl},
         year = 2008,
        month = apr,
       volume = {677},
       number = {1},
        pages = {L63},
          doi = {10.1086/587838},
       adsurl = {https://ui.adsabs.harvard.edu/abs/2008ApJ...677L..63M}
}

@ARTICLE{2008ApJ...684L..55M,
       author = {{Moreno}, F. and {Ortiz}, J.~L. and {Santos-Sanz}, P. and {Morales}, N. and {Vidal-N{\'u}{\~n}ez}, M.~J. and {Lara}, L.~M. and {Guti{\'e}rrez}, P.~J.},
        title = "{Erratum: ``A Model of the Early Evolution of the 2007 Outburst of Comet 17P/Holmes'' (ApJ, 677, L63 [2008])}",
      journal = {\apjl},
         year = 2008,
        month = sep,
       volume = {684},
       number = {1},
        pages = {L55},
          doi = {10.1086/592043},
       adsurl = {https://ui.adsabs.harvard.edu/abs/2008ApJ...684L..55M}
}

@ARTICLE{2022NatSR..12.5855L,
       author = {{Lecavelier des Etangs}, Alain and {Cros}, Lucie and {H{\'e}brard}, Guillaume and {Martioli}, Eder and {Duquesnoy}, Marc and {Kenworthy}, Matthew A. and {Kiefer}, Flavien and {Lacour}, Sylvestre and {Lagrange}, Anne-Marie and {Meunier}, Nad{\`e}ge and {Vidal-Madjar}, Alfred},
        title = "{Exocomets size distribution in the {\ensuremath{\beta}} Pictoris planetary system}",
      journal = {Scientific Reports},
         year = 2022,
        month = apr,
       volume = {12},
          eid = {5855},
        pages = {5855},
          doi = {10.1038/s41598-022-09021-2},
archivePrefix = {arXiv},
       eprint = {2204.13618},
 primaryClass = {astro-ph.EP},
       adsurl = {https://ui.adsabs.harvard.edu/abs/2022NatSR..12.5855L}
}

@ARTICLE{1981A&A....98...45W,
       author = {{Wallis}, M.~K. and {MacPherson}, A.~K.},
        title = "{On the outgassing and jet thrust of snowball comets}",
      journal = {\aap},
         year = 1981,
        month = may,
       volume = {98},
       number = {1},
        pages = {45-49},
       adsurl = {https://ui.adsabs.harvard.edu/abs/1981A&A....98...45W}
}

@ARTICLE{2019A&A...625L..13Z,
       author = {{Zieba}, S. and {Zwintz}, K. and {Kenworthy}, M.~A. and {Kennedy}, G.~M.},
        title = "{Transiting exocomets detected in broadband light by TESS in the {\ensuremath{\beta}} Pictoris system}",
      journal = {\aap},
         year = 2019,
        month = may,
       volume = {625},
          eid = {L13},
        pages = {L13},
          doi = {10.1051/0004-6361/201935552},
archivePrefix = {arXiv},
       eprint = {1903.11071},
 primaryClass = {astro-ph.SR},
       adsurl = {https://ui.adsabs.harvard.edu/abs/2019A&A...625L..13Z}
}

@ARTICLE{1999A&AS..140...15L,
       author = {{Lecavelier Des Etangs}, A.},
        title = "{A library of stellar light variations due to extra-solar comets}",
      journal = {\aaps},
         year = 1999,
        month = nov,
       volume = {140},
        pages = {15-20},
          doi = {10.1051/aas:1999114},
       adsurl = {https://ui.adsabs.harvard.edu/abs/1999A&AS..140...15L}
}

@ARTICLE{1968P&SS...16.1221W,
       author = {{Wallis}, M.~K.},
        title = "{The fountain model for plasma envelopes of comets}",
      journal = {\planss},
         year = 1968,
        month = oct,
       volume = {16},
       number = {10},
        pages = {1221-1248},
          doi = {10.1016/0032-0633(68)90028-7},
       adsurl = {https://ui.adsabs.harvard.edu/abs/1968P&SS...16.1221W}
}

@ARTICLE{1957Tell....9...92A,
       author = {{Alfven}, H.},
        title = "{On the theory of comet tails.}",
      journal = {Tellus},
         year = 1957,
        month = jan,
       volume = {9},
        pages = {92},
          doi = {10.3402/tellusa.v9i1.9064},
       adsurl = {https://ui.adsabs.harvard.edu/abs/1957Tell....9...92A}
}

@ARTICLE{1999A&A...343..916L,
       author = {{Lecavelier Des Etangs}, A. and {Vidal-Madjar}, A. and {Ferlet}, R.},
        title = "{Photometric stellar variation due to extra-solar comets}",
      journal = {\aap},
         year = 1999,
        month = mar,
       volume = {343},
        pages = {916-922},
          doi = {10.48550/arXiv.astro-ph/9812381},
archivePrefix = {arXiv},
       eprint = {astro-ph/9812381},
 primaryClass = {astro-ph},
       adsurl = {https://ui.adsabs.harvard.edu/abs/1999A&A...343..916L}
}

@INPROCEEDINGS{2008LPI....39.1627T,
       author = {{Trigo-Rodriguez}, J.~M. and {Davidsson}, B. and {Montanes-Rodriguez}, P. and {Sanchez}, A. and {Troughton}, B.},
        title = "{All-Sky Cameras Detection and Telescope Follow-Up of the 17P/Holmes Outburst}",
    booktitle = {39th Annual Lunar and Planetary Science Conference},
         year = 2008,
       series = {Lunar and Planetary Science Conference},
        month = mar,
        pages = {1627},
       adsurl = {https://ui.adsabs.harvard.edu/abs/2008LPI....39.1627T}
}

@MISC{optool,
       author = {{Dominik}, Carsten and {Min}, Michiel and {Tazaki}, Ryo},
        title = "{OpTool: Command-line driven tool for creating complex dust opacities}",
 howpublished = {Astrophysics Source Code Library, record ascl:2104.010},
         year = 2021,
        month = apr,
          eid = {ascl:2104.010},
        pages = {ascl:2104.010},
archivePrefix = {ascl},
       eprint = {2104.010},
       adsurl = {https://ui.adsabs.harvard.edu/abs/2021ascl.soft04010D}
}

@ARTICLE{Draine2003,
       author = {{Draine}, B.~T.},
        title = "{Scattering by Interstellar Dust Grains. I. Optical and Ultraviolet}",
      journal = {\apj},
         year = 2003,
        month = dec,
       volume = {598},
       number = {2},
        pages = {1017-1025},
          doi = {10.1086/379118},
archivePrefix = {arXiv},
       eprint = {astro-ph/0304060},
 primaryClass = {astro-ph},
       adsurl = {https://ui.adsabs.harvard.edu/abs/2003ApJ...598.1017D}
}

@ARTICLE{Henning1996,
       author = {{Henning}, T. and {Stognienko}, R.},
        title = "{Dust opacities for protoplanetary accretion disks: influence of dust aggregates.}",
      journal = {\aap},
         year = 1996,
        month = jul,
       volume = {311},
        pages = {291-303},
       adsurl = {https://ui.adsabs.harvard.edu/abs/1996A&A...311..291H}
}

@ARTICLE{Dorschner1995,
       author = {{Dorschner}, J. and {Begemann}, B. and {Henning}, T. and {Jaeger}, C. and {Mutschke}, H.},
        title = "{Steps toward interstellar silicate mineralogy. II. Study of Mg-Fe-silicate glasses of variable composition.}",
      journal = {\aap},
         year = 1995,
        month = aug,
       volume = {300},
        pages = {503},
       adsurl = {https://ui.adsabs.harvard.edu/abs/1995A&A...300..503D}
}

@ARTICLE{Zubko1996,
       author = {{Zubko}, V.~G. and {Mennella}, V. and {Colangeli}, L. and {Bussoletti}, E.},
        title = "{Optical constants of cosmic carbon analogue grains - I. Simulation of clustering by a modified continuous distribution of ellipsoids}",
      journal = {\mnras},
         year = 1996,
        month = oct,
       volume = {282},
       number = {4},
        pages = {1321-1329},
          doi = {10.1093/mnras/282.4.1321},
       adsurl = {https://ui.adsabs.harvard.edu/abs/1996MNRAS.282.1321Z}
}

@ARTICLE{Reach2009,
       author = {{Reach}, William T. and {Vaubaillon}, Jeremie and {Kelley}, Michael S. and {Lisse}, Carey M. and {Sykes}, Mark V.},
        title = "{Distribution and properties of fragments and debris from the split Comet 73P/Schwassmann-Wachmann 3 as revealed by Spitzer Space Telescope}",
      journal = {\icarus},
         year = 2009,
        month = oct,
       volume = {203},
       number = {2},
        pages = {571-588},
          doi = {10.1016/j.icarus.2009.05.027},
archivePrefix = {arXiv},
       eprint = {0905.3162},
 primaryClass = {astro-ph.EP},
       adsurl = {https://ui.adsabs.harvard.edu/abs/2009Icar..203..571R}
}

@software{radmc,
       author = {{Dullemond}, C.~P. and {Juhasz}, A. and {Pohl}, A. and {Sereshti}, F. and {Shetty}, R. and {Peters}, T. and {Commercon}, B. and {Flock}, M.},
        title = "{RADMC-3D: A multi-purpose radiative transfer tool}",
 howpublished = {Astrophysics Source Code Library, record ascl:1202.015},
         year = 2012,
        month = feb,
          eid = {ascl:1202.015},
archivePrefix = {ascl},
       eprint = {1202.015},
       adsurl = {https://ui.adsabs.harvard.edu/abs/2012ascl.soft02015D}
}

@ARTICLE{Kral2018,
       author = {{Kral}, Quentin and {Wyatt}, Mark C. and {Triaud}, Amaury H.~M.~J. and {Marino}, Sebastian and {Th{\'e}bault}, Philippe and {Shorttle}, Oliver},
        title = "{Cometary impactors on the TRAPPIST-1 planets can destroy all planetary atmospheres and rebuild secondary atmospheres on planets f, g, and h}",
      journal = {\mnras},
         year = 2018,
        month = sep,
       volume = {479},
       number = {2},
        pages = {2649-2672},
          doi = {10.1093/mnras/sty1677},
archivePrefix = {arXiv},
       eprint = {1802.05034},
 primaryClass = {astro-ph.EP},
       adsurl = {https://ui.adsabs.harvard.edu/abs/2018MNRAS.479.2649K}
}

@ARTICLE{Strom2020,
       author = {{Str{\o}m}, Paul A. and {Bodewits}, Dennis and {Knight}, Matthew M. and {Kiefer}, Flavien and {Jones}, Geraint H. and {Kral}, Quentin and {Matr{\`a}}, Luca and {Bodman}, Eva and {Capria}, Maria Teresa and {Cleeves}, Ilsedore and {Fitzsimmons}, Alan and {Haghighipour}, Nader and {Harrison}, John H.~D. and {Iglesias}, Daniela and {Kama}, Mihkel and {Linnartz}, Harold and {Majumdar}, Liton and {de Mooij}, Ernst J.~W. and {Milam}, Stefanie N. and {Opitom}, Cyrielle and {Rebollido}, Isabel and {Rogers}, Laura K. and {Snodgrass}, Colin and {Sousa-Silva}, Clara and {Xu}, Siyi and {Lin}, Zhong-Yi and {Zieba}, Sebastian},
        title = "{Exocomets from a Solar System Perspective}",
      journal = {\pasp},
         year = 2020,
        month = oct,
       volume = {132},
       number = {1016},
          eid = {101001},
        pages = {101001},
          doi = {10.1088/1538-3873/aba6a0},
archivePrefix = {arXiv},
       eprint = {2007.09155},
 primaryClass = {astro-ph.EP},
       adsurl = {https://ui.adsabs.harvard.edu/abs/2020PASP..132j1001S}
}

@ARTICLE{Ferlet1987,
       author = {{Ferlet}, R. and {Hobbs}, L.~M. and {Vidal-Madjar}, A.},
        title = "{The beta Pictoris circumstellar disk. V. Time variations of the Ca II-K line.}",
      journal = {\aap},
         year = 1987,
        month = oct,
       volume = {185},
        pages = {267-270},
       adsurl = {https://ui.adsabs.harvard.edu/abs/1987A&A...185..267F}
}

@ARTICLE{Kiefer2014,
       author = {{Kiefer}, F. and {Lecavelier des Etangs}, A. and {Boissier}, J. and {Vidal-Madjar}, A. and {Beust}, H. and {Lagrange}, A.-M. and {H{\'e}brard}, G. and {Ferlet}, R.},
        title = "{Two families of exocomets in the {\ensuremath{\beta}} Pictoris system}",
      journal = {\nat},
         year = 2014,
        month = oct,
       volume = {514},
       number = {7523},
        pages = {462-464},
          doi = {10.1038/nature13849},
       adsurl = {https://ui.adsabs.harvard.edu/abs/2014Natur.514..462K}
}

@ARTICLE{Iglesias2025,
       author = {{Iglesias}, Daniela and {Rebollido}, Isabel and {Norazman}, Azib and {Snodgrass}, Colin and {Seligman}, Darryl Z. and {Xu}, Siyi and {Hoeijmakers}, H. Jens and {Kenworthy}, Matthew and {Lecavelier des Etangs}, Alain and {Bannister}, Michele and {Yang}, Bin},
        title = "{An Overview of Exocomets}",
      journal = {\ssr},
         year = 2025,
        month = dec,
       volume = {221},
       number = {8},
          eid = {122},
        pages = {122},
          doi = {10.1007/s11214-025-01247-6},
archivePrefix = {arXiv},
       eprint = {2511.08270},
 primaryClass = {astro-ph.EP},
       adsurl = {https://ui.adsabs.harvard.edu/abs/2025SSRv..221..122I}
}

@ARTICLE{Rappaport2018,
       author = {{Rappaport}, S. and {Vanderburg}, A. and {Jacobs}, T. and {LaCourse}, D. and {Jenkins}, J. and {Kraus}, A. and {Rizzuto}, A. and {Latham}, D.~W. and {Bieryla}, A. and {Lazarevic}, M. and {Schmitt}, A.},
        title = "{Likely transiting exocomets detected by Kepler}",
      journal = {\mnras},
         year = 2018,
        month = feb,
       volume = {474},
       number = {2},
        pages = {1453-1468},
          doi = {10.1093/mnras/stx2735},
archivePrefix = {arXiv},
       eprint = {1708.06069},
 primaryClass = {astro-ph.EP},
       adsurl = {https://ui.adsabs.harvard.edu/abs/2018MNRAS.474.1453R}
}

@INPROCEEDINGS{Tinetti2022,
       author = {{Tinetti}, Giovanna and {Eccleston}, Paul and {Lueftinger}, Theresa and {Salvignol}, Jean-Christophe and {Fahmy}, Salma and {Alves de Oliveira}, Caterina},
        title = "{Ariel: Enabling planetary science across light-years}",
    booktitle = {European Planetary Science Congress},
         year = 2022,
        month = sep,
          eid = {EPSC2022-1114},
        pages = {EPSC2022-1114},
          doi = {10.5194/epsc2022-1114},
archivePrefix = {arXiv},
       eprint = {2104.04824},
 primaryClass = {astro-ph.IM},
       adsurl = {https://ui.adsabs.harvard.edu/abs/2022EPSC...16.1114T}
}

@dataset{2024yCat..19330054L,
       author = {{Lu}, C.~X. and {Chen}, C.~H. and {Sargent}, B.~A. and {Watson}, D.~M. and {Lisse}, C.~M. and {Green}, J.~D. and {Sitko}, M.~L. and {Mittal}, T. and {Lebouteiller}, V. and {Sloan}, G.~C. and {Rebollido}, I. and {Hines}, D.~C. and {Girard}, J.~H. and {Werner}, M.~W. and {Stapelfeldt}, K.~R. and {Wu}, W. and {Worthen}, K.},
        title = "{VizieR Online Data Catalog: IRTF/SpeX and Spitzer/IRS spectra of beta Pictoris (Lu+, 2022)}",
 howpublished = {VizieR On-line Data Catalog: J/ApJ/933/54. Originally published in: 2022ApJ...933...54L},
         year = 2024,
        month = apr,
          eid = {J/ApJ/933/54},
       adsurl = {https://ui.adsabs.harvard.edu/abs/2024yCat..19330054L}
}

@ARTICLE{2022A&A...660A..49P,
       author = {{Pavlenko}, Ya. and {Kulyk}, I. and {Shubina}, O. and {Vasylenko}, M. and {Dobrycheva}, D. and {Korsun}, P.},
        title = "{New exocomets of {\ensuremath{\beta}} Pic}",
      journal = {\aap},
         year = 2022,
        month = apr,
       volume = {660},
          eid = {A49},
        pages = {A49},
          doi = {10.1051/0004-6361/202142111},
archivePrefix = {arXiv},
       eprint = {2202.13373},
 primaryClass = {astro-ph.SR},
       adsurl = {https://ui.adsabs.harvard.edu/abs/2022A&A...660A..49P}
}

@ARTICLE{2020ExA....50..303M,
       author = {{Mugnai}, Lorenzo V. and {Pascale}, Enzo and {Edwards}, Billy and {Papageorgiou}, Andreas and {Sarkar}, Subhajit},
        title = "{ArielRad: the Ariel radiometric model}",
      journal = {Experimental Astronomy},
         year = 2020,
        month = oct,
       volume = {50},
       number = {2-3},
        pages = {303-328},
          doi = {10.1007/s10686-020-09676-7},
archivePrefix = {arXiv},
       eprint = {2009.07824},
 primaryClass = {astro-ph.IM},
       adsurl = {https://ui.adsabs.harvard.edu/abs/2020ExA....50..303M}
}

@techreport{TESS_SDPDD_2020,
    title = {TESS Science Data Products Description Document},
    author = {{TESS Science Processing Operations Center}},
    number = {EXP-TESS-ARC-ICD-0014, Rev. F},
    year = {2020},
    month = {September},
    institution = {NASA Ames Research Center},
    url = {https://heasarc.gsfc.nasa.gov/docs/tess/data-products.html}
}

@ARTICLE{2025MNRAS.542.1486N,
       author = {{Norazman}, Azib and {Kennedy}, Grant M. and {Cody}, Ann Marie and {Giles}, Daniel and {Gill}, Samuel and {Kruse}, Ethan},
        title = "{A search for transiting exocomets in TESS sectors 1-26}",
      journal = {\mnras},
         year = 2025,
        month = sep,
       volume = {542},
       number = {2},
        pages = {1486-1508},
          doi = {10.1093/mnras/staf1298},
archivePrefix = {arXiv},
       eprint = {2508.04673},
 primaryClass = {astro-ph.EP},
       adsurl = {https://ui.adsabs.harvard.edu/abs/2025MNRAS.542.1486N}
}

@ARTICLE{2025A&A...704A.191D,
       author = {{Dumond}, P. and {Lecavelier des Etangs}, A. and {Kiefer}, F. and {H{\'e}brard}, G. and {Caill{\'e}}, V.},
        title = "{Search for exocomet transits in Kepler light curves: Ten new transits identified}",
      journal = {\aap},
         year = 2025,
        month = dec,
       volume = {704},
          eid = {A191},
        pages = {A191},
          doi = {10.1051/0004-6361/202556033},
archivePrefix = {arXiv},
       eprint = {2510.14687},
 primaryClass = {astro-ph.EP},
       adsurl = {https://ui.adsabs.harvard.edu/abs/2025A&A...704A.191D}
}

@ARTICLE{2026A&A...707A..60V,
       author = {{Vrignaud}, T. and {Lecavelier des Etangs}, A.},
        title = "{Transit distances and compositions of low-velocity exocomets in the {\ensuremath{\beta}} Pic system}",
      journal = {\aap},
         year = 2026,
        month = mar,
       volume = {707},
          eid = {A60},
        pages = {A60},
          doi = {10.1051/0004-6361/202557819},
archivePrefix = {arXiv},
       eprint = {2601.15216},
 primaryClass = {astro-ph.EP},
       adsurl = {https://ui.adsabs.harvard.edu/abs/2026A&A...707A..60V}
}

@ARTICLE{2025A&A...697A..21V,
       author = {{Vrignaud}, T. and {Lecavelier des Etangs}, A. and {Str{\o}m}, P.~A. and {Kiefer}, F.},
        title = "{Abundances of refractory ions in Beta Pictoris exocomets}",
      journal = {\aap},
         year = 2025,
        month = may,
       volume = {697},
          eid = {A21},
        pages = {A21},
          doi = {10.1051/0004-6361/202453568},
archivePrefix = {arXiv},
       eprint = {2503.17346},
 primaryClass = {astro-ph.EP},
       adsurl = {https://ui.adsabs.harvard.edu/abs/2025A&A...697A..21V}
}

@ARTICLE{2019A&A...630A..26M,
       author = {{Mannel}, T. and {Bentley}, M.~S. and {Boakes}, P.~D. and {Jeszenszky}, H. and {Ehrenfreund}, P. and {Engrand}, C. and {Koeberl}, C. and {Levasseur-Regourd}, A.~C. and {Romstedt}, J. and {Schmied}, R. and {Torkar}, K. and {Weber}, I.},
        title = "{Dust of comet 67P/Churyumov-Gerasimenko collected by Rosetta/MIDAS: classification and extension to the nanometer scale}",
      journal = {\aap},
         year = 2019,
        month = oct,
       volume = {630},
          eid = {A26},
        pages = {A26},
          doi = {10.1051/0004-6361/201834851},
       adsurl = {https://ui.adsabs.harvard.edu/abs/2019A&A...630A..26M}
}









\appendix

\section{Stellar limb-darkening}\label{app:limb-darkening}

Stellar limb-darkening coefficients were calculated for the same
solar-like stellar parameters adopted in the radiative-transfer
simulations, namely $T_{\rm eff}=5780$\,K, $\log g=4.5$, and solar
metallicity. We used ExoTETHyS \citep{2020JOSS....5.1834M,
2020AJ....159...75M} with PHOENIX stellar-atmosphere models
\citep{2012A&A...546A..14C,2013A&A...552A..16C,2013A&A...553A...6H}
and the four-parameter Claret limb-darkening law
\citep{2000A&A...363.1081C}. For the Ariel passbands, we relied on the precomputed grid published by \cite{2022ExA....53..533M}.

\begin{table}
	\centering
	\caption{Four-parameter Claret limb-darkening coefficients derived from the PHOENIX stellar-atmosphere models for the synthetic photometric bands. The table lists the corresponding band and coefficients from $a_1$ to $a_4$.}
	\label{tab:LDCs_plot_PHOENIX}
	\begin{tabular}{l|rrrr} 
		\hline
		Band                         &   a$_1$      &     a$_2$      &     a$_3$      &      a$_4$\\
                                     &              &                &                &            \\
		\hline
    	VISPhot                      & 0.42035764   &   0.00537437   &   0.80322100   &  -0.43604644\\
		FGS1                         & 0.55864516   &  -0.15417819   &   0.63826082   &  -0.32330794\\
		FGS2                        & 0.55782921   &  -0.15544589   &   0.41624407   &  -0.20822700\\
        J$_{\mathrm{synth}}$         & 0.42589018   &   0.20599435   &  -0.03546119   &  -0.05474250\\
        H$_{\mathrm{synth}}$          & 0.52090889   &   0.29601097   &  -0.45146255   &   0.14425193\\
        K$_{\mathrm{short,synth}}$   & 0.58725190   &  -0.23321094   &   0.16168500   &  -0.07159309\\
        K$_{\mathrm{long,synth}}$   & 0.58081412   &  -0.38066047   &   0.29853821   &  -0.10349102\\
        IR1$_{\mathrm{synth}}$      & 0.40605806   &  -0.15098702   &   0.06995198   &  -0.01444635\\
        IR2$_{\mathrm{synth}}$       & 0.39579374   &  -0.23162147   &   0.15119499   &  -0.04301039\\
        IR3$_{\mathrm{synth}}$       & 0.34804203   &  -0.25051555   &   0.18807609   &  -0.05782028\\
        TESS                         & 0.56249908   &  -0.16231451   &   0.55765752   &  -0.27956643\\
        CHEOPS                        & 0.48196366   &  -0.09090522   &   0.74537715   &  -0.38632087\\
		\hline
	\end{tabular}
\end{table}


\bsp	
\label{lastpage}
\end{document}